# Feedback control logic synthesis for non safe Petri nets

Dideban A*. Alla H.**

*Semnan University, IRAN (Tel: (+98)231-3354123;  e-mail: adideban@ Semnan.ac.ir).
** GIPSA lab, 38402 St Martin d'Heres Cedex FRANCE , Hassane.alla@inpg.fr)

**Abstract** – This paper addresses the problem of forbidden states of non safe Petri Net (PN) modelling discrete events systems. To prevent the forbidden states, it is possible to use conditions or predicates associated with transitions. Generally, there are many forbidden states, thus many complex conditions are associated with the transitions. A new idea for computing predicates in non safe Petri nets will be presented. Using this method, we can construct a maximally permissive controller if it exists.

*Keywords*: Controller Synthesis, Discrete event system, Forbidden states, Petri Nets

## 1. INTRODUCTION

Real – life discrete-event systems (DES) are becoming more and more complex and highly automated which makes it tricky the realization of an efficient and realistic control system. Given a discrete-event model of the plant and the specification of the desired behaviour, the objective is to synthesize appropriate supervisor that will act in closed-loop with the plant according to the desired behaviour. Finite-state machines and formal languages are the modelling framework considered in the approach of Ramadge and Wonham (1989). The main limitation in such an approach is the lack of structure in controlled automata.

Petri nets have been proposed as an alternative modelling formalism for DES control. There have been many attempts to solve the control problem for DES with PN modelling. Li and Wonham (1994) have presented an algorithm, which calculates the optimal solution for nets whose uncontrollable subnets are loop-free. The theory of regions (Ghaffari et al. 2003a), allows the design of a maximally permissive PN controller. However, the number of control places is equal to the number of forbidden states and sometimes this leads to complex solutions. Holloway and Krogh have presented a method for controller calculating in real time for a safe and cyclic marked graph (Holloway and Krogh,1990). An effective method for controller synthesis was presented in (Dideban and Alla, 2006), however this method is applicable only on safe PNs. Moreover the final model may be complex.

In this paper, a method is presented to solve the problem of forbidden states for controlled Petri Nets. We develop the method presented in (Dideban and Alla, 2006) for **non safe PNs**. Moreover, in comparison with (Ghaffari et al. 2003b) and (Holloway et al. 1996),, the final condition will be very simple. The disadvantage of this approach is the calculation of the reachability graph that is fortunately performed off-line. In this paper we use the "*over-state*" concept that was presented in (Dideban and Alla, 2008).

This paper is organized as follows: In the second section, the fundamental definitions will be presented. The motivations for this approach will be presented in Section 3 and in Section 4, a method for calculating the condition of forbidding transitions will be presented. Then, in Section *5*, the method for simplification of the conditions in safe PNs is called. In Section 6, a compact algorithm will formalize this method and solving the problem of forbidden states will be illustrated via an example. In Section 7, this method will be extended for non safe PN. Finally, the conclusion is given in the last section.

## 2. FUNDEMENTAL DEFINITIONS

*2.1 .Petri Nets*

A PN is presented by a 4-uple $N = \{P, T, W, C\}$ where: 1) $P$ is the set of places, 2) $T$ is the set of transitions, 3) $W$: $(P \times T) \cup (P \times T)$, is the incidence matrix, and 4) $C$ is the firing conditions associated with each controllable transition.

The reachability graph consists of nodes, which correspond to the accessible markings $M_i$, and arcs to the firing of the transitions. In the reachability graph, there are two types of states: the authorized state $M_A$ and the forbidden state $M_F$. Among the forbidden states, a particular and important subset is constituted by the border forbidden states, which are denoted by the set $M_B$. These states are such that all the input transitions are controllable.

In this paper, we use the word state instead of marking.

*Definition* 1**:** The set $\{0,1\}^N$ represents all the Boolean vectors of dimension $N$.   ❑

The set of the marked places of a marking $M$ is given by a Support function that is defined in the following.

*Definition* 2: The function *Support*($X$) of a vector $X \in \{0,1\}^N$ is:  *Support*(X) = The set of marked places in vector X.   ❑

The support of vector $M_0^T = [1, 0, 1, 0, 0, 1, 0]$ is:

Support $(M_0) = \{P_1 P_3 P_6\}$

*Definition* 3: Let $M_1$ and $M_2$ be 2 states of the system, and $P = \{P_1, P_2, ..., P_N\}$ the PN set of places, $M_2$ is an over-state of $M_1$ if: $\forall P_i \in P / m_1(P_i) \geq m_2(P_i)$ and $\exists P_i \in P / m_1(P_i) > m_2(P_i)$
This relation is represented as shown bellow:
$$M_1 > M_2 \qquad \square$$

*Definition* 4: Informally, the forbidden states are:
- The states reachable in the process but not authorized by the specification.
- Deadlock states.
$\square$

## 2.2. Critical and sound transitions

In the PN modelling, when a controllable event is associated with a transition, a controller can be calculated for this transition. Then we use the controllable transition instead of the controllable event. Firing of some controllable transitions can lead to forbidden states. This set is named s*et of critical transitions*. The rest of the controllable transition is named *sound transitions*.

## 2.3. Critical and sound states

The border forbidden states are reached from the admissible states by the occurrence of controllable events. Preventing the occurrence of the controllable events can forbid entering to a forbidden state. By constructing the reachability graph, we can divide the admissible states for each controllable transition ($t_i$) into 3 groups:

- The states from which the firing of $t_i$ is possible and allowed;

- The states from which the firing of $t_i$ is possible but is not allowed.

- The states from which the firing of $t_i$ is not possible;

The first group is named sound states and corresponds to the states from which by firing transition $t_i$, the admissible states can be reached.

The second group of these states corresponds to the states leading to a border forbidden state by firing $t_i$. This group is named critical states.

The third group are the states for which the firing of this transition is not possible. The first and third groups are non critical states. The first and second group can be defined as below;

*Definition 5*: Let $M_B$ be the set of border forbidden states and $M_A$ the set of admissible states. The sets of $t_i$ critical states $M_{t_i}^C$, and $t_i$ sound states $M_{t_i}^S$, are defined as follows:

$$\forall t_i \in \Sigma_c \quad M_{t_i}^C = \{M_i \in M_A \mid M_i \xrightarrow{t_i} M_j, M_j \in M_B\}$$

$$M_{t_i}^S = \{M_i \in M_A \mid M_i \xrightarrow{t_i} M_j, M_j \in M_A\}$$

Where $\Sigma_c$ is the set of controllable transitions $\square$

*Definition 6*: Let $M_{t_i}^C$ be the set of critical states for the critical transition $t_i$. Control $Ut_i$: $(M_{t_i}^C, M_j) \to \{0,1\}$ is defined as follow:

$$U_{t_i}(M_{t_i}^C, M_j) = \begin{cases} 0 & M_j \in M_{t_i}^C \\ 1 & \text{if not} \end{cases} \qquad \square$$

The control relation is modelled in Figure 1.

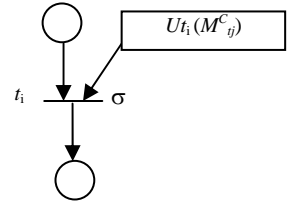

Fig. 1. Adding the condition as a control

This is similar with the approach presented in (Holloway et al. 1996). The difference between both approaches is the method of calculation of the control $U_{ti}$. As it will be shown the advantage of our approach is to provide a method to determine simple forbidding conditions. To achieve this goal, we need to build the reachability graph as an intermediate step. Our approach is applicable to ordinary PNs. Firstly we present it on safe PN and then for non safe PN. It is supposed that all of the events are independent.

## 3. MOTIVATION

We, first present our ideas via a simple example. Consider the classical system composed of two machines $M_1$ and $M_2$ and one buffer $S_1$. The specification constraint is the capacity of the buffer (Figure 2). Firstly we suppose that the capacity of $S_1$ is one part, it will be changed later in order to have a non safe model.

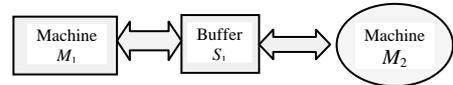

Fig. 2. A simple system

We suppose here that only the starts of the tasks (event $c_1$, $c_2$, i.e. transitions $t_1$, $t_3$) are controllable and the ends of task (event $f_1$, $f_2$, i.e. transitions $t_2$, $t_4$) are uncontrollable. The desired functioning in closed loop for this system is given in Figure 3.

The goal is to find a control such that the border forbidden states are never reached. For this, we must construct the rechability graph of the closed loop model. The rechability graph for this example is given in Figure 4.

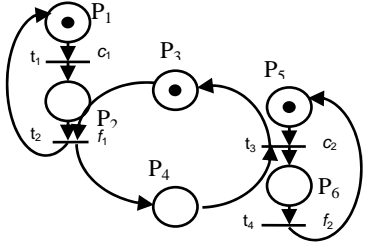

Fig. 3. Desired functioning in closed loop

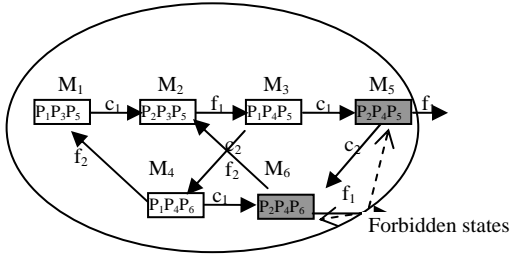

Fig. 4. Reachability graph

So we must control transition $t_1$ (event $c_1$) in states $M_3$ and $M_4$, therefore:

$$M_{t_1}^C = \{M_3, M_4\}$$

In the behaviour of this PN, some transitions associated with uncontrollable events lead to forbidden states. For example, the firing of $t_2$ is possible when place $P_2$ is marked and event $f_1$ occurs even place $P_3$ is empty. These states are called the forbidden states and correspond to the set of border forbidden states for this example: $M_B = M_F = \{P_2P_4P_5, P_2P_4P_6\}$

We can compute the forbidden states by the method that is presented in (Kumar and Holloway 1996).

This can be accomplished by adding conditions to transition $t_1$ resulting to the transition to be blocked when the system is in states $M_3$ and $M_4$. This condition can be computed for each state $M_j$ taking into account the presence of marks in the places: $Ut_1(M_{t_1}^C) = ((m(P_1) \wedge m(P_4) \wedge m(P_5)) \vee (m(P_1) \wedge m(P_4) \wedge m(P_6)))'$

*Remark*2: Variable $m_j(P_i)$ represents the number of marks in place $P_i$ in state $M_j$ and then for a safe PN is a Boolean variable. Moreover the condition $Ut_1(\ )$ is also Boolean. ❏

Logical expression $Ut_1(\ )$ means that transition $t_1$ would not be fireable when the system is in states $M_3$ or $M_4$. The situation in $M_3$ is presented in Figure 5.

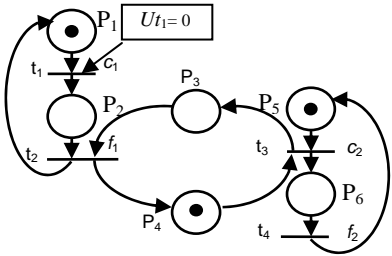

Fig. 5. Adding the condition as a control

In this state the firing of $t_1$ is not possible. Now, the generalization of this idea is given in the following section and a method will be presented for the simplification of these conditions thanks to the concepts of over-state which corresponds to a significant contribution.

## 4. CALCULATION OF FORBIDDING CONDITIONS

From the sets $M_\sigma^C$ and $M_\sigma^S$ for each controllable event or transition, there are two ways to construct the controller: the calculation of the conditions of forbidding or the calculation of the conditions for enabling each controllable event. In this paper, the first method is called but in general case it is better to examine both methods and to select the simpler solution. Now we explain how these states can be forbidden (Dideban and Alla, 2006).

*Property* 1: Let $M_1$ be a critical state for transition $t_i$. By using the control $Ut_i(M_1)$, we can forbid the firing of $t_i$ in state $M_1$.
$$U_{t_i}(M_1) = (\prod_{r=1}^{card(\sup port(M_1))} m(P_{1r}))'$$ ❏

By this method, it is possible to forbid the firing from state $M_1$, but other states can forbid the firing of $t_i$ by the same control. For example if there is a sound state $M_2$ such that $M_2 < M_1$ ($M_2$ is an over-state of $M_1$), this state also will be considered as a critical sate. Then in this case the controller is not maximal permissive. For constructing a maximal permissive controller, the modification of the conditions will be considered in the next property.

*Property* 2: Let $M_{t_i}^C = M_1, M_2, \ldots, M_m$, be the set of all the $t_i$ critical- states, the condition for forbidding the firing towards the forbidden states will be calculated as follows:

$$U_{t_i}(M_{t_i}^C) = (\bigcup_{k=1}^{card(M_{t_i}^C)} (\prod_{r=1}^{card(\sup port(M_k))} m(P_{kr})))'$$ ❏

Now, the condition of firing can be calculated for each transition. However, sometimes the conditions are complex and makes very difficult to understand the dynamic behaviour of the system. Moreover, the calculation time for the conditions in real time can be very large. It is then necessary to use similar simplification methods presented in (Dideban and Alla, 2005, 2008).

## 5. SIMPLIFICATION OF THE CONDITIONS

The simplification method that is presented in this paper is not exactly the same that presented in (Dideban and Alla, 2008). It is based on the concepts of critical and sound states which are reachable sates, while the previous approach uses mainly the concepts of forbidden states.

*Property* 3: Let $M_2$ be an over-state of a $t_i$ critical state $M_1$. Forbidding the firing of transition $t_i$ by control $Ut_i(M_2)$ leads to forbidding the firing of transition $t_i$ in state $M_1$.

$Ut_i(M_2) = 0$ and $M_2 < M_1 \Rightarrow$

Transition $t_i$ is non fireable from $M_1$ ❏

This property guaranties that transition $t_i$ is not fireable in state $M_1$ by using of any over-state of $M_1$ in $t_i$ control. But the condition deduced from an over-state may forbid the firing of transition $t_i$ in some of the $t_i$ sound states. This problem is formalized via the following property:

*Property* 4: Let $M_{ti}^C$ be the set of $t_i$ critical states and $M_{ti}^S$ the set of $t_i$ sound states such that $M_1 \in M_{ti}^C$ and $M_2 < M_1$.

The control $Ut_i(M_1)$ can be replaced by the control $Ut_i(M_2)$ if there is no state $M_3 \in M_{ti}^S$ such that: $M_2 < M_3$ ❑

An over-state can often cover several critical states, and then the condition computed for this over-state gives the control for all these critical states. Then the forbidding condition will be simpler. Our objective is to find the over-states that cover the maximum number of critical states. To reach this goal, a set of all over-states for the $t_i$ critical-state is constructed and the $t_i$ sound states should be deleted from this set. This construction is similar to the one presented in (Dideban and Alla, 2008).

## 6. CONTROLLER SYNTHESIS

Firstly the different steps for the controller synthesis are presented by an algorithm and then illustrated via an example.

*Algorithm* 1:

1: Calculate the set of border forbidden states $M_B$.

2: Calculate the set of the critical-states and the transitions that can be fired (C).

3: Reorganize set C for each transition ($M_{t_i}^C$).

4: Calculate set of sound-states($M_{t_i}^S$) for all transitions in C.

5: Calculate the forbidding conditions for each controllable transition.

6: Simplify the forbidding conditions (Property 4)

7: Select the necessary and sufficient set of over-states.

*Safe PN example*: For the example presented in Section 3, the following results are obtained:

$Ut_1(M_{t_1}^C) = ((m(P_1) \wedge m(P_4) \wedge m(P_5)) \vee (m(P_1) \wedge m(P_4) \wedge m(P_6)))`$

In the following, the simplification method is described.

In the first step in Section 2, the set of critical states for each controllable transition were calculated.

$$M_{t_1}^C = \{P_1P_4P_5, P_1P_4P_6\}$$

Now we calculate the set of sound states for each controllable transition. The occurring of event $c_1$ (transition $t_1$) is only authorized in state $M_1$ (Figure 5).

$$M_{t_1}^S = \{P_1P_3P_5\}$$

Using the same method presented in (Dideban and Alla, 2008) for each controllable transition, the sets of all over-states for critical states and sound states must be constructed. For transition $t_1$, the set of its over-states for $M_{t_1}^C$, ($C_1^{t_1}$) and $M_{t_1}^S$ ($S_1^{t_1}$) are computed as follow:

$C_1^{t_1} = \{P_1, P_4, P_5, P_1P_4, P_1P_5, P_4P_5, P_1P_4P_5, P_6, P_1P_6, P_4P_6, P_1P_4P_6\}$

$S_1^{t_1} = \{P_1, P_3, P_5, P_1P_3, P_1P_5, P_3P_5, P_1P_3P_5\}$

Now, all over-states that exist in set $S_1^{t_1}$ should be deleted from set $C_1^{t_1}$.

$C_2^{t_1} = C_1^{t_1} \setminus S_1^{t_1} = \{P_4, P_1P_4, P_4P_5, P_1P_4P_5, P_6, P_1P_6, P_4P_6, P_1P_4P_6\}$

In addition, the over-states which are covered by another over-state can be deleted. $C_3^{t_1} = \{P_4, P_6\}$

Now there are two over-states for transition $t_1$. The simpler condition for each transition must be selected. These conditions must forbid the firing of these transitions in all critical states (Table 1).

Table 1. Final choice for $t_1$ in safe PN

| Critical states  $C_3^{t_1}$ | $P_1P_4P_5$ | $P_1P_4P_6$ | Choice($C_4^{t_1}$) |
|---|---|---|---|
| $P_4$ | √ | √ | √ |
| $P_6$ |   | √ |   |
|   | √ | √ |   |

For selecting the simpler condition, we write all of the critical states in first line and all of the simplified over–states in first column. For each over–state, we mark all of the states covered by it. Now for choosing the final result as for the Cluskey method (Morris Mano 2001), firstly we select the over–state covering only one state. Then for the state that is covered by two or more over–states, we select the over–state that covers the more non selected states. The final condition for our example is: $C_4^{t_1} = \{P_4\} \Rightarrow Ut_1(M_{t_1}^C) = (m(P_4))`$

Controllable transition $t_3$ has not to be controlled since no forbidden state is reached.

There is a dual method for the controller synthesis. We can calculate the firing conditions instead of forbidding conditions. For this we must change the critical and sound sets. In this case for our example we have:

$S_2^{t_1} = S_1^{t_1} \setminus C_1^{t_1} = \{P_3, P_3P_5, P_1P_3P_5\}$

$S_3^{t_1} = \{P_3\}$  $S_4^{t_1} = \{P_3\}$  $Ut_1(M_{t_1}^C) = (m(P_3))$

The simpler solution must be kept. Here they are equivalent.

## 7. EXTENSION OF THE SIMPLIFICATION METHOD TO NON SAFE PETRI NETS

In non safe Petri Nets, the number of marks in a place is not a Boolean variable. In this section, the controller synthesis method will be developed for non safe Petri Nets.

## 7.1 New Definition

*Definition* 7: The function *Support(X)* of a vector $X \in \{0,1,\ldots,i,\ldots; i \in \mathbb{N}\}^N$ is presented as below:

*Support*(X) = The set of marked places in vector X with the number of marks presented as the power for each place. ❑

The support of vector $M_0^T = [1, 0, 2, 0, 0, 1, 0]$ is:

Support $(M_0) = \{P_1 P_3^2 P_6\} = \{P_1 P_3 P_3 P_6\}$

There is no change in the definition of an over-state. If $M_2$ is an over-state of $M_1$, then $M_2 < M_1$

Let $M_{t_i}^C$ be the set of critical states for a controllable transition $t_i$. The control $Ut_i$: $(M_{t_i}^C, M_j) \to \{0,1\}$ is defined exactly in the same way:

$$U_{t_i}(M_{t_i}^C, M_j) = \begin{cases} 0 & M_j \in M_{t_i}^C \\ 1 & \text{if not} \end{cases}$$

We have seen that the prevention of firing in the critical state in a safe PN is simple to compute. For example, for the critical state $P_1 P_3 P_6$ we can use the Boolean condition $m_j(P_1) \cdot m_j(P_3) \cdot m_j(P_6)$. But how is that for non safe PN? A definition is presented as bellow:

*Definition* 8: Let $M_1 = (P_{11}^{k1} P_{12}^{k2} \ldots P_{1m}^{km} \ldots P_{1n}^{kn})$ be a critical state where $P_{1m}$ represents the marked place, $km^1$ the number of marks and $Ut_i(M_1)$, the condition for preventing of firing from $M_1$, $Ut_i(M_1)$ can be calculated as this:

$U_{ti}(M_1) = (C_{11} C_{12} \ldots C_{1m} \ldots C_{1n})'$

So that $C_{1m} = \begin{cases} 1 & m_j(P_{1m}) \geq k_{1m} \\ 0 & m_j(P_{1m}) < k_{1m} \end{cases}$ ❑

The simplification method by using the concept of over-state is different of the one used for safe PNs. But the definition of an over-state is not changed for a non safe PN.

*Definition* 9: In non safe PNs, a state $M_2 = (P_{21}^{k21} P_{22}^{k22} \ldots P_{2m}^{k2m} \ldots P_{2n}^{k2n})$ is an over-state of a state $M_1 = (P_{11}^{k11} P_{12}^{k12} \ldots P_{1m}^{k1m} \ldots P_{1n}^{k1n})$ if for $\forall P_{2i} \in support(M_2) \; \exists P_{1i} \in support(M_1) / P_{1i} == P_{2i}$ and $k_{2i} \leq k_{1i}$ then:

$$M_2 < M_1 \qquad ❑$$

However, there is a need of a new definition for the Boolean control given by an over-state. Since this control must cover its original state, it is equal to one (or satisfied) if the number of marks in each place is greater or equal to the number of corresponding marks in the over-state. This new definition for the control calculation is presented as bellow:

*Definition* 10: Let $M_1 = (P_{11}^{k1} P_{12}^{k2} \ldots P_{1m}^{km} \ldots P_{1n}^{kn})$ be a critical state or an over-state deduced from it, that $P_{1m}$ present the marked places and $km$ present the number of marks and

---
[1] By convention, if $m(P_{km}) = 0$, we say that $km = 0$..

$Ut_i(M_1)$, the condition for preventing of firing in $M_j$, $Ut_i(M_1)$ can be calculated as this:

$U_{ti}(M_1) = (C_{11} C_{12} \ldots C_{1m} \ldots C_{1n})'$

So that $C_{1m} = \begin{cases} 1 & m_j(P_{1m}) \geq k_{1m} \\ 0 & m_j(P_{1m}) < k_{1m} \end{cases}$

❑

As for safe PNs, this property guaranties that transition $t_i$ is not fireable from any over-state of $M_1$. And by testing the sound states, we are sure that no authorized state is forbidden.

*Remark* 3: By this definition, it is possible to arrive to an empty set after the reduction process. In this case we can use Definition 8 for the control calculation which cannot be simplified.

## 7.2 Example of a non safe PN

Consider the example presented in Figure 3 modified such that the capacity of buffer is 2. The closed loop PN model for this the system is given in Figure 6.

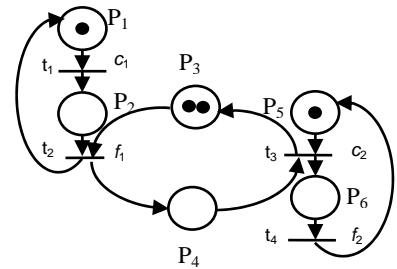

Fig. 6. Desired functioning in closed loop in a non safe PN model

The marking graph for this example is given in Figure 7.

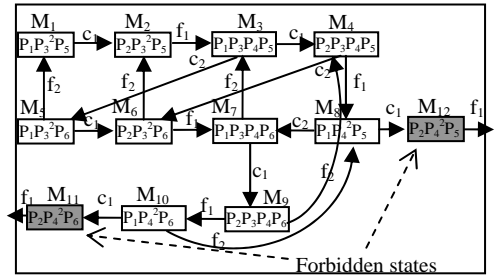

Fig. 7. Marking graph for a non safe PN

As previously, the goal is to find a control such that the border forbidden states are never reached. So we must control transition $t_1$ (event $c_1$) in states $M_8$ and $M_{10}$, therefore:

$M_{t_1}^C = \{M_8, M_{10}\}$   $M_8 = P_1 P_4^2 P_5$ , $M_{10} = P_1 P_4^2 P_6$

$Ut_1(M_{t_1}^C) = ((C_{81} \wedge C_{82} \wedge C_{83}) \vee (C_{101} \wedge C_{102} \wedge C_{103}))`$

$C_{81} := m(P_1) \geq 1$ , $C_{82} := (m(P_4) \geq 2)$ , $C_{83} := m(P_5) \geq 1$

$C_{101} := m(P_1) \geq 1$ , $C_{102} := (m(P_4) \geq 2)$ , $C_{103} := m(P_6) \geq 1$

This condition will be simplified using algorithm 1.

For each controllable transition we must calculate the sets of critical states $M_{t_1}^C$ and sound states $M_{t_1}^S$.

$M_{t_1}^C = \{P_1P_4^2P_5, P_1P_4^2P_6\}$

The set of sound states for $t_1$ is:

$M_{t_1}^S = \{P_1P_3^2P_5, P_1P_3P_4P_5, P_1P_3^2P_6, P_1P_3P_4P_6\}$

The set of over-states for critical states of $t_1$ is:

$C_1^{t1} = \{P_1, P_4, P_5, P_1P_4, P_1P_5, P_4^2, P_4P_5, P_1P_4^2, P_4^2P_5, P_1P_4P_5, P_1P_4^2P_5, P_6, P_1P_6, P_4^2P_6, P_1P_4^2P_6\}$

And for sound states:

$S_1^{t1} = \{P_1, P_3, P_4, P_5, P_6, P_1P_3, P_1P_5, P_3P_5, P_3^2, P_1P_3^2, P_3^2P_5, P_1P_3P_5, P_1P_3^2P_5, P_1P_4, P_4P_5, P_1P_3P_4, P_1P_4P_5, P_3P_4P_5, P_1P_3P_4P_5, P_1P_6, P_3P_6, P_3^2P_6, P_1P_3P_6, P_1P_3^2P_6, P_4P_6, P_3P_4P_6, P_1P_3P_4P_6\}$

Now we calculate the set of over-states of critical states that are not sound states.

$C_2^{t1} = C_1^{t1} \setminus S_1^{t1} = \{P_4^2, P_1P_4^2, P_4^2P_5, P_1P_4^2P_5, P_4^2P_6, P_1P_4^2P_6\}$

$C_3^{t1} = \{P_4^2\}$

The final choice is the same that used for safe Petri Nets model. $C_4^{t1} = \{P_4^2\} \Rightarrow Ut_1(M_{t_1}^C) := (m(P_4) \geq 2)$'

**Table 2. Final choice for $t_1$ in non safe PN**

| Critical states $C_3^{t1}$ | $P_1P_4^2P_5$ | $P_1P_4^2P_6$ | Choice ($C_4^{t1}$) |
|---|---|---|---|
| $P_4^2$ | √ | √ | √ |
| | √ | √ | |

That means if the number of marks in place $P_4$ equal to 2, this condition equal to zero and transition $t_1$ is not fireable. The controlled model is presented in Figure 8.

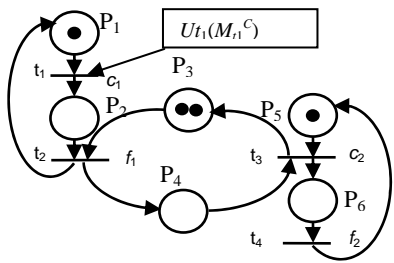

Fig. 8. Final controlled model

## 8. CONCLUSION

This paper presents an efficient approach for solving forbidden states control problems. Petri Nets are used for modelling discrete event systems and the controls correspond to conditions associated with transitions. The basic idea is to use simpler conditions for preventing the forbidden states. A condition is a Boolean expression deduced from the marking of special places for every controllable event. The concept of over-state covering several markings allows significant simplification of these conditions, which are determined in a formal way. This concept has been generalized to non safe Petri nets taking into account the number of marks in each place.

This method needs the determination of the reachability graph, hopefully it is done offline and the size of the final model is close to the specifications. It is maximal permissive, since all of the used properties meet this condition. By using a dual method sometimes it is possible to calculate a simpler controller. Each one gives a maximal permissive controller and the simplest model must be chosen.